\documentclass[12pt]{article}
\pdfoutput=1
\usepackage{putex}
\usepackage{feyn}
\usepackage[vcentermath]{youngtab}
\usepackage{subfig}
\usepackage{lscape}

\usepackage{graphicx}
\usepackage{epstopdf}
\usepackage{enumerate}
\usepackage{cite}
\usepackage{tensor}
\usepackage{slashed}
\usepackage{amsmath}
\usepackage{amssymb}
\usepackage{mathrsfs}
\usepackage{lgrind}

\usepackage{hyperref}

\numberwithin{equation}{section}

\newcommand {\be} {\begin {equation}}
\newcommand {\ee} {\end {equation}}

\newcommand {\bes} {\begin {equation*}}
\newcommand {\ees} {\end {equation*}}

\newcommand{\Par}{\partial_{\mu}}

\newcommand{\eps}{\epsilon}

\def\CH{{\cal H}}

\newcommand{\beq}{\begin{equation}}
\newcommand{\eeq}{\end{equation}}

\def\be{ \begin{equation} }
\def\ee{ \end{equation} }

\begin{document}

\preprint{PUPT-2478}

\institution{PU}{Department of Physics, Princeton University, Princeton, NJ 08544}
\institution{PCTS}{Princeton Center for Theoretical Science, Princeton University, Princeton, NJ 08544}

\title{
Critical $Sp(N)$ Models in $6-\eps$ Dimensions and Higher Spin dS/CFT
}

\authors{Lin Fei,\worksat{\PU} Simone Giombi,\worksat{\PU} Igor R.~Klebanov\worksat{\PU,\PCTS} and Grigory Tarnopolsky\worksat{\PU}
}

\abstract{Theories of anti-commuting scalar fields are non-unitary, but they are of interest both in statistical mechanics
and in studies of the higher spin de Sitter/Conformal Field Theory correspondence. We consider an $Sp(N)$ invariant theory of
$N$ anti-commuting scalars and one commuting scalar, which has cubic interactions and is renormalizable in 6 dimensions.
For any even $N$ we find an IR stable fixed point in $6-\epsilon$ dimensions at imaginary values of coupling constants. Using calculations up to three loop order, we develop $\epsilon$ expansions for several
operator dimensions and for the sphere free energy $F$. The conjectured $F$-theorem is obeyed in spite of the non-unitarity of the theory.
The $1/N$ expansion in the $Sp(N)$ theory is related to that in the corresponding
$O(N)$ symmetric theory by the change of sign of $N$. Our results point to the existence of interacting non-unitary 5-dimensional CFTs with $Sp(N)$ symmetry,
where operator dimensions are real. We conjecture that these CFTs are dual to the minimal higher spin theory in 6-dimensional de Sitter space
with Neumann future boundary conditions on the scalar field.
For $N=2$ we show that the IR fixed point possesses an enhanced global symmetry
given by the supergroup $OSp(1|2)$. This suggests the existence of $OSp(1|2)$ symmetric CFTs in dimensions smaller than 6.
We show that the $6-\eps$ expansions of the scaling dimensions and sphere free energy in our $OSp(1|2)$ model are the same as in the
 $q \rightarrow 0$ limit of the $q$-state Potts model.}

\date{}
\maketitle

\tableofcontents

\section{Introduction and Summary}

Among the classic models of quantum field theory, a prominent role is played by the $O(N)$ invariant theories of $N$ massless scalar fields $\phi^i$,
which interact via the potential ${\lambda \over 4} (\phi^i \phi^i)^2$. For any positive $N$ these models possess interacting IR fixed points in dimensions
$2<d<4$ \cite{Wilson:1971dc}.
These theories contain $O(N)$ singlet current operators with all even spin, and when $N$ is large the current anomalous dimensions are $\sim 1/N$
\cite{Wilson:1973jj}. Since
all the higher spin currents are nearly conserved, the $O(N)$ models possess a weakly broken higher spin symmetry. In the
Anti-de Sitter/Conformal Field Theory (AdS/CFT) correspondence \cite{Maldacena:1997re,Gubser:1998bc,Witten:1998qj},
each spin $s$ conserved current in a $d$ dimensional CFT is mapped to a massless
spin $s$ gauged field in $d+1$ dimensional AdS space. For these reasons, it was conjectured \cite{Klebanov:2002ja} that the singlet sector of
the critical $O(N)$ models in $d=3$ is
dual to the interacting higher spin theory in $d=4$ containing massless gauge fields of all even positive spin
\cite{Fradkin:1987ks,Vasiliev:1990en,Vasiliev:1992av, Vasiliev:2003ev}.
This minimal Vasiliev theory also contains a scalar field with $m^2=-2/\ell^2_{AdS}$, and the two admissible boundary conditions on this field
\cite{Breitenlohner:1982jf,Klebanov:1999tb}
distinguish the interacting $O(N)$ model from the free
one (in the latter all the higher spin currents are conserved exactly). A review of the higher spin AdS/CFT dualities may be be found in
\cite{Giombi:2012ms}.

A remarkable feature of the Vasiliev theories
\cite{Fradkin:1987ks,Vasiliev:1990en,Vasiliev:1992av, Vasiliev:2003ev}
is that they are consistent not only in Anti-de Sitter, but also in de Sitter space.
On general grounds one expects a CFT dual to quantum gravity in $dS_4$ to be a non-unitary theory defined in three dimensional Euclidean space \cite{Strominger:2001pn}.
In \cite{Anninos:2011ui} it was proposed  that the CFT dual to the minimal higher spin theory in $dS_4$ is the theory of an even number $N$ of anti-commuting scalar fields $\chi^i$ with the action
\be
S=\int d^3 x \bigg (
\frac{1}{2} \Omega_{ij} \partial_\mu \chi^i \partial^\mu \chi^j +\frac{\lambda}{4} (\Omega_{ij}\chi^i\chi^j)^2
\bigg )
\label{symplectic}
\ .\ee
This theory possesses $Sp(N)$ symmetry, and $\Omega_{ij}$ is the invariant symplectic matrix. This model was originally introduced and studied in
\cite{LeClair:2006kb,LeClair:2007iy},
where it was shown to possess an IR fixed point in $4-\epsilon$ dimensions. The beta function of this model is related to that of the $O(N)$ model via the
replacement $N\rightarrow -N$. According to the proposal of \cite{Anninos:2011ui}, the free UV fixed point of (\ref{symplectic}) is dual
to the minimal higher spin theory in $dS_4$ with the Neumann future boundary conditions on the $m^2=2/\ell^2_{dS}$ bulk scalar field, and its interacting IR fixed point
to the same higher spin theory but with the Dirichlet boundary conditions on the scalar field. In the latter case, the higher spin symmetry
is slightly broken at large $N$, and the de Sitter higher spin gauge fields are expected to acquire small masses through quantum effects.
\footnote{A potential difficulty with this picture is that unitarity in $dS_{d+1}$ space requires that a massive field of spin $s>1$ should satisfy $m^2 > (s-1)(s+d-3)/\ell^2_{dS}$ \cite{Higuchi:1986py, Higuchi:1986wu, Deser:2001us}. In other words, there is a finite gap between massive fields and massless
ones (the latter are dual to exactly conserved currents in the CFT). However, since the masses are generated by quantum effects and are parametrically small at large $N$,
this is perhaps not fatal for bulk unitarity. It would be interesting to clarify this further.} A discussion of the de Sitter boundary conditions from the point of view
of the wave function of the Universe was given in \cite{Maldacena:2002vr,Anninos:2012ft}.

In this paper we consider an extension of the proposed higher spin dS/CFT correspondence \cite{Anninos:2011ui} to higher dimensional de Sitter spaces,
and in particular to $dS_6$.
Our construction mirrors our recent work \cite{Fei:2014yja,Fei:2014xta} on the higher dimensional extensions of the higher spin AdS/CFT.
It was observed long ago  \cite{Parisi:1975im,parisi1977non,Bekaert:2011cu} that in $d>4$ the quartic $O(N)$ models possess UV fixed points which
can be studied in the large $N$ expansion. The UV completion of the $O(N)$ scalar theory in $4<d<6$ was proposed in
\cite{Fei:2014yja}; it is the cubic $O(N)$ symmetric theory of $N+1$ scalar fields
$\sigma$ and $\phi^i$:
\be
S=\int d^d x \bigg (\frac{1}{2}(\Par \phi^i)^2 + \frac{1}{2}(\Par \sigma)^2+ \frac{1}{2}g_1 \sigma (\phi^i)^2 + \frac{1}{6}g_2 \sigma^3\bigg)
\ .
\label{onmodels}
\ee
For sufficiently large $N$, this theory has an IR stable fixed point in $6-\epsilon$ dimensions with real values of $g_1$ and $g_2$.
The beta functions and anomalous dimensions were calculated to three loop order \cite{Fei:2014yja,Fei:2014xta},\footnote{The one loop beta functions of the model
(\ref{onmodels}) in $d=6$ were first calculated in \cite{Ma:1975vn}.} and the available results agree nicely with the $1/N$ expansion
of the quartic $O(N)$ model at its UV fixed point
\cite{Vasiliev:1981yc,Vasiliev:1982dc,Lang:1990ni, Lang:1992zw, Petkou:1994ad}
when the quartic $O(N)$ model is continued to $6-\epsilon$ dimensions. The conformal bootstrap approach to the higher dimensional $O(N)$ model was explored in
\cite{Nakayama:2014yia,Chester:2014gqa,Bae:2014hia}.

To extend the idea of \cite{Anninos:2011ui} to $dS_{d+1}$ with $d>4$, we may consider non-unitary CFTs (\ref{symplectic}) which in $d>4$ possess UV fixed points
for large $N$. The $1/N$ expansion of operator scaling dimensions may be developed using the generalized Hubbard-Stratonovich transformation, and one finds
that it is related to the $1/N$ expansion in the $O(N)$ models via the replacement $N\rightarrow -N$. In $d=5$ this interacting fixed point
should be dual to the higher spin theory in $dS_6$ with Neumann boundary conditions on the $m^2=6/\ell^2_{dS}$ scalar field (corresponding to the
conformal dimension $\Delta=2+\mathcal{O}(1/N)$ on the CFT side).\footnote{The $d=5$ free $Sp(N)$ model corresponds to Dirichlet scalar
boundary conditions in the $dS_6$ dual. It should be possible to extend this free $Sp(N)$ model/$dS$ higher spin duality to general dimensions,
since the Vasiliev equations in $(A)dS_{d+1}$ are known for all $d$ \cite{Vasiliev:2003ev}.}
In search of the UV completion of these quartic CFTs in $4<d<6$, we introduce
the cubic theory of one commuting real scalar field $\sigma$ and
$N$ anti-commuting scalar fields $\chi^i$:
\be
S=\int d^d x \bigg (
\frac{1}{2} \Omega_{ij} \partial_\mu \chi^i \partial^\mu \chi^j +\frac{1}{2}\left(\partial_{\mu}\sigma\right)^2+ \frac{1}{2}g_1 \Omega_{ij}\chi^i\chi^j\,\sigma+ \frac{1}{6}g_2 \sigma^3
\bigg )
\label{cubicsymplectic}
\ .\ee
Alternatively, we may combine the fields $\chi^i$ into $N/2$ complex anti-commuting scalars $\theta^\alpha$, $\alpha=1,\ldots, N/2$
\cite{LeClair:2006kb,LeClair:2007iy}; then the action assumes the form
\be
S=\int d^d x \bigg (
\partial_\mu \theta^\alpha \partial^\mu \bar \theta^\alpha +\frac{1}{2}\left(\partial_{\mu}\sigma\right)^2+ g_1 \sigma \theta^\alpha \bar \theta^\alpha + \frac{1}{6}g_2 \sigma^3
\bigg )
\label{complexsymplectic}
\ .\ee
We study the beta functions for this theory in $d=6-\epsilon$ and show that they are related to the beta functions of the theory (\ref{onmodels}) via the replacement
$N\rightarrow -N$. For all $N$ there exists an IR fixed point of the theory (\ref{cubicsymplectic}) with imaginary values of $g_1$ and $g_2$. This
is similar to the IR fixed point of the single scalar cubic field theory (corresponding to the $N=0$ case of our models), which was used by Fisher \cite{Fisher:1978pf} as an approach to the Lee-Yang edge singularity. The fact that the couplings are purely imaginary makes the integrand of the path integral
oscillate rapidly at large $\sigma$; this should be contrasted with real couplings giving a potential unbounded from below.

Our results allow us to study the $6-\epsilon$ expansion of the theory (\ref{cubicsymplectic}) with arbtrary $N$,
and we observe that at finite $N$ there are qualitative differences between the $Sp(N)$ and $O(N)$ models which are not
seen in the $1/N$ expansion.
In fact, for the $Sp(N)$ model  there is no analogue of the lower bound $N_{crit}$ that
was found in the $O(N)$ case \cite{Fei:2014yja}. For the lowest value, $N=2$, we observe some special phenomena.
In this theory, which contains two real anti-commuting scalars, it is impossible to formulate the quartic interaction (\ref{symplectic}); thus, the cubic lagrangian
(\ref{cubicOSp}) seems to be the only possible description of the interacting theory with global $Sp(2)$ symmetry. Furthermore, it
becomes enhanced to the supergroup
$OSp(1|2)$ because at the IR fixed point the two coupling constants are related via $g_2^*= 2 g_1^*$.
The enhanced symmetry implies that the scaling dimensions of $\sigma$ and $\chi^i$ are equal, and we check this to order $\epsilon^3$.
An example of theory with $OSp(1|2)$ symmetry is provided by the $q\rightarrow 0$ limit of the $q$-state Potts model
\cite{Caracciolo:2004hz}.\footnote{We are grateful to Giorgio Parisi for informing us about this and for important discussions.}
We show that the $6-\eps$ expansions of the scaling dimensions in our $OSp(1|2)$ symmetric theory are the same as in the
$q\rightarrow 0$ limit of the $q$-state Potts model.\footnote{We thank Sergio Caracciolo for suggesting this comparison to us and for informing us about the paper \cite{2007PhRvL..98c0602D}.} 
This provides strong evidence that the $OSp(1|2)$ symmetric IR fixed point of the cubic theory (\ref{cubicOSp}) describes the second order transitions in the ferromagnetic $q=0$ Potts model, which exist in $2<d<6$ \cite{2007PhRvL..98c0602D}.

Using the results of \cite{Giombi:2014xxa}, we also compute perturbatively the sphere free energies of the models (\ref{cubicsymplectic}).
The sphere free energy for the $OSp(1|2)$ symmetric model is found to be the same as for the $q=0$ Potts model.
In terms of the quantity $\tilde F = \sin(\frac{\pi d}{2})\log Z_{S^d}$, which was introduced in \cite{Giombi:2014xxa}
as a natural way to generalize the $F$-theorem \cite{Casini:2011kv, Jafferis:2011zi, Klebanov:2011gs} to continuous dimensions, we find that the RG flow in the cubic $Sp(N)$ models in $d=6-\epsilon$
satisfies ${\tilde F}_{\textrm{UV}}>{\tilde F}_{\textrm{IR}}$ for all $N\ge 2$. We show that the same result holds in the model (\ref{symplectic}) in $d=4-\epsilon$.
\footnote{The models (\ref{symplectic}) and (\ref{cubicsymplectic}) also satisfy the $F$-theorem to leading order in the large $N$ expansion (for all $d$),
since the leading order correction to $\tilde {F}_{\textrm{UV}}-\tilde{F}_{\textrm{IR}}$ is of order $N^0$, and was shown to satisfy the $F$-theorem
in the corresponding unitary $O(N)$ models \cite{Klebanov:2011gs,Giombi:2014iua, Giombi:2014xxa}.}
This is somewhat surprising, since for non-unitary CFTs the inequality ${\tilde F}_{\textrm{UV}}>{\tilde F}_{\textrm{IR}}$ is not always satisfied. It would be interesting
to understand if this is related to the ``pseudo-unitary" structure discussed in \cite{LeClair:2007iy}, and to the fact that these models are presumably
dual to unitary higher spin gravity theories in de Sitter space.



\section{The IR fixed points of the cubic $Sp(N)$ theory}

The beta functions and anomalous dimensions for the $Sp(N)$ symmetric model (\ref{cubicsymplectic}) can be obtained by replacing $N\rightarrow -N$ in the corresponding
results for the cubic $O(N)$ model (\ref{onmodels}), which were computed in
\cite{Fei:2014yja,Fei:2014xta} to three loop order. Indeed, writing the action in the complex basis (\ref{complexsymplectic}), we see that the Feynman
rules and propagators are identical to those of the $O(N)$ theory written in the $U(N/2)$ basis, the only difference being that the $N/2$
complex scalars are anticommuting. Hence, for each closed loop of the $\theta^{\alpha}$ we get an extra minus sign, thus explaining the
replacement $N\rightarrow -N$.

Using the results in \cite{Fei:2014yja,Fei:2014xta}, the beta functions for the $Sp(N)$ model are then found to be:
\begin{equation}
\begin{aligned}
\beta_{1}=&-\frac{\epsilon}{2}g_{1} -\frac{1}{12(4\pi)^{3}} g_{1}  \left((N+8)g_1^2+12g_1 g_2-g_2^2 \right)\\
&-\frac{1}{432(4\pi)^{6}} g_{1}  \left((536-86N)g_1^4+12(30+11N)g_1^3 g_2+(628-11N)g_1^2 g_2^2+24 g_1 g_2^3-13 g_2^4 \right)+\ldots, \\
\beta_{2}=&-\frac{\epsilon}{2}g_{2} +\frac{1}{4(4\pi)^{3}} \left(4Ng_1^3-Ng_1^2 g_2 -3g_2^3 \right)\\
&+\frac{1}{144(4\pi)^{6}} \left( 24Ng_1^5+322N g_1^4 g_2 + 60N g_1^3g_2^2-31N g_1^2 g_2^3-125g_2^5 \right)+\ldots.\\
\end{aligned}
\label{beta1}
\end{equation}
We have omitted the explicit three loop terms, which can be obtained from \cite{Fei:2014yja,Fei:2014xta}.
Similarly, the anomalous dimensions of the fields $\chi^i$ and $\sigma$ take the form
\begin{equation}
\begin{aligned}
\gamma_{\chi}=&\frac{g_1^2}{6(4\pi)^3} +  \frac{g_1^2}{432(4\pi)^{6}}  \left(g_1^2 (11 N+26)+48 g_1 g_2 -11 g_2^2\right)+\ldots, \\
\gamma_{\sigma} =&-\frac{Ng_1^2-g_2^2}{12(4\pi)^3} -\frac{1}{432(4\pi)^{6}} \left(2N g_1^4 +48 N  g_1^3 g_2-11 N g_1^2 g_2^2-13 g_2^4\right)+\ldots. \\
\end{aligned}
\label{dimsig}
\end{equation}

With the beta functions at hand, we can look for non-trivial fixed points of the RG flow satisfying $\beta_1(g_1^*,g_2^*)=0, \beta_2(g_1^*,g_2^*)=0$. For all positive $N$, we find two physically equivalent fixed points with purely imaginary coupling constants, hence all operator dimensions remain real. These fixed points are IR stable for all $N$ (the stability matrix $M_{ij}=\frac{\partial \beta_i}{\partial g_j}$ has positive eigenvalues). Note that this is different from the $O(N)$ versions of these models \cite{Fei:2014yja,Fei:2014xta}, where one finds a critical $N$, whose one loop value is $\simeq 1038$, below which the IR stable fixed points with real coupling constants disappear. However, to all orders in the $1/N$ expansion, the fixed point couplings and conformal dimensions in the $Sp(N)$ models are related to the ones in the $O(N)$ models by the replacement $N\rightarrow -N$.

\begin{figure}[t]
\centering
\includegraphics[width=10cm]{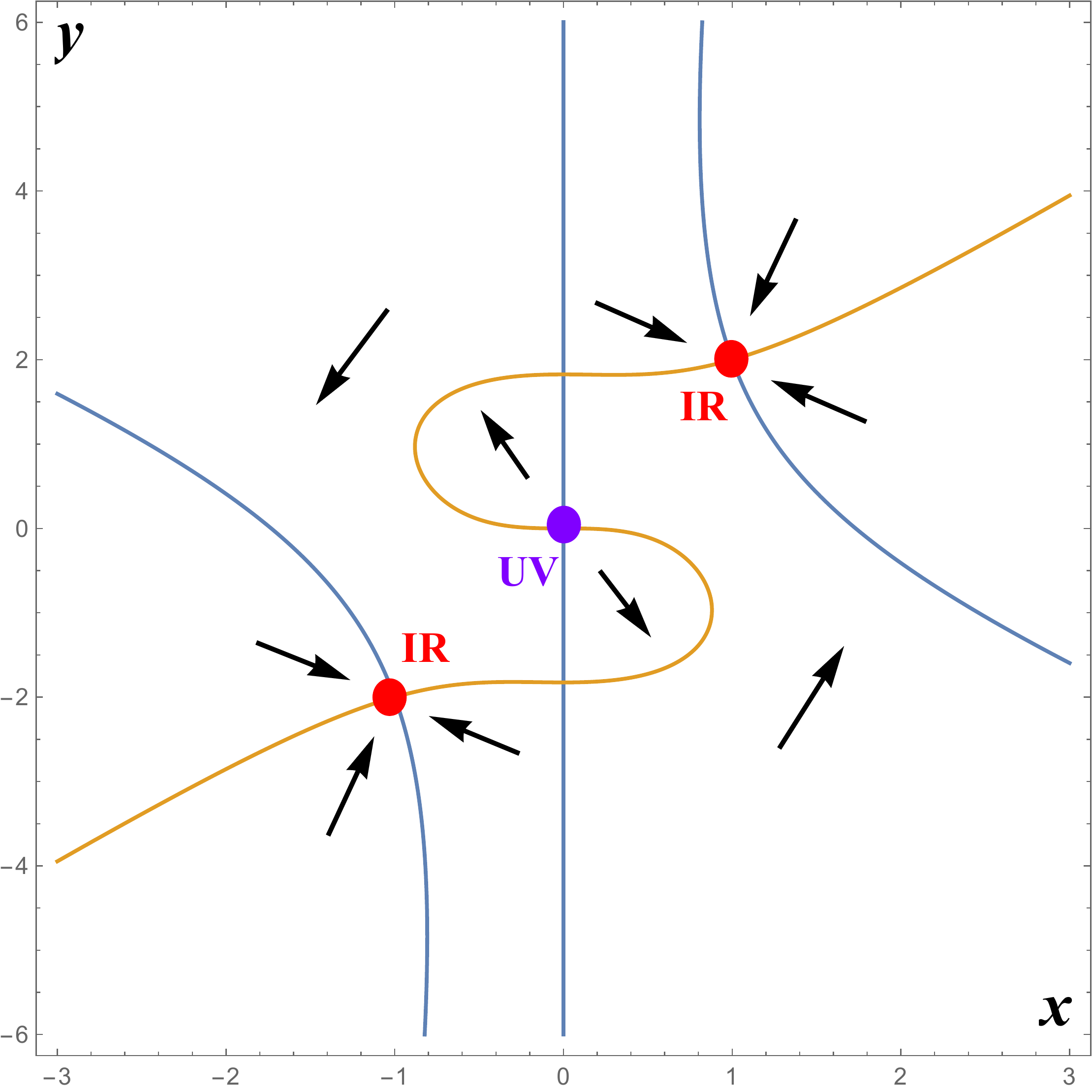}
\caption{The zeroes of the one loop $\beta$ functions and the RG flow directions for the $OSp(1|2)$ model.
The coordinates are defined via $g_1 = i\sqrt{\frac{(4\pi)^3\epsilon}{5}} x$, $g_2 = i\sqrt{\frac{(4\pi)^3\epsilon}{5}} y$, and
the red dots correspond to the stable IR fixed points.}
\label{Sp2_flow}
\end{figure}

Figure \ref{Sp2_flow} shows the RG flow directions for $N=2$.
The arrows indicate how the coupling constants flow towards the IR. The two IR fixed points are physically equivalent because they are related by $g_i\rightarrow -g_i$. At higher values of $N$, the qualitative behavior of the RG flows and fixed points remain the same. We still have a UV Gaussian fixed point, and two stable IR fixed points.

A special structure emerges for $N=2$. In this case
we find the fixed point solution
\begin{equation}
g_2^*=2 g_1^*\,,\qquad g_1^* = i\sqrt{\frac{(4\pi)^3\epsilon}{5}}\left(1+\frac{67}{180}\epsilon+\mathcal{O}(\epsilon^2)\right)\ ,
\label{N2-fixed}
\end{equation}
and the conformal dimensions of the fundamental fields are equal (the three loop term given below can be obtained from the results in \cite{Fei:2014yja,Fei:2014xta})
\begin{equation}
\Delta_{\sigma}=\Delta_{\chi} = 2 - \frac{8}{15}\epsilon - \frac{7}{450} \epsilon^2-\frac{269 - 702 \zeta(3)}{33750} \epsilon^3+\ldots\,.
\label{N2-dim}
\end{equation}
We show in the next section that the equality of dimensions is a consequence of a symmetry enhancement from $Sp(2)$ to the supergroup $OSp(1|2)$.

It is natural to ask whether symmetry enhancement can occur at other values of $N$. For instance, we can explicitly check for which $N$ the dimensions of $\sigma$ and $\chi$ are equal. A direct calculation using the beta functions and anomalous dimensions up to three loops shows that this only happens for $N=2$ and $N=-1$. The latter case corresponds
to the 3-state Potts model fixed point of the the theory with two commuting scalars \cite{Fei:2014xta}: it has $g_1^*=-g_2^*$ and enhanced $Z_3$ symmetry.\footnote{For $N=-1$, there is also a (non-unitary) solution with $g_1^*=g_2^*$ which corresponds to two decoupled Fisher models \cite{Fisher:1978pf}, and hence the dimensions of the fundamental fields are trivially equal.}

\begin{figure}[h!!]
\centering
\includegraphics[width=16cm]{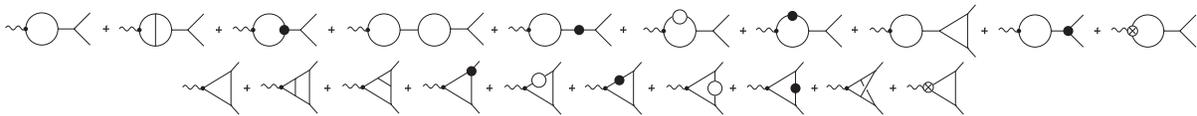}
\caption{Diagrams contributing to the mixing of $\sigma^2$ and $\Omega_{ij}\chi^i\chi^j$ operators to two loop order.}
\label{diagrams}
\end{figure}

The anomalous dimensions of some composite operators may be similarly obtained from the results in \cite{Fei:2014yja,Fei:2014xta}. Let us quote the explicit result for
the quadratic operators arising from the mixture of $\sigma^2$ and $\Omega_{ij}\chi^i\chi^j$. These operators have the same classical dimension,
so we expect them to mix. The $2\times 2$ anomalous dimension mixing matrix $\gamma^{ab}$ was given in \cite{Fei:2014yja,Fei:2014xta} up to one-loop order.
Extending those results to two loops, we find the mixing matrix
\begin{align}
\!\!\!\!\!\!\!\!\!\!
\left( \begin{array}{cc}
    -\frac{g_{1}^2 (N+4)}{6 (4 \pi )^{3}}+ \frac{g_{1}^2 \left(g_{1}^2 (21 N-134)-6 g_{1}g_{2} (2 N+5)+5 g_{2}^2\right)}{108 (4 \pi )^6} &\frac{Ng_{1}  (6 g_{1}-g_{2})}{6 (4 \pi )^{3}} + \frac{N g_{1}^2  \left(18g_{1}^2+161 g_{1} g_{2}+15 g_{2}^2\right)}{108 (4 \pi )^6}\\
    \frac{g_{1} (g_{2}-6 g_{1})}{6 (4 \pi )^{3}} -\frac{g_{1} \left(6 g_{1}^3 (11 N+20)-g_{1}^2 g_{2} (11 N-324)+12 g_{1} g_{2}^2-13 g_{2}^3\right)}{216 (4 \pi )^6}&
-\frac{N g_{1}^2+4 g_{2}^2}{6 (4 \pi )^{3}} +\frac{Ng_{1}^{2}(160  g_{1}^2 +36 g_{1} g_{2} -41   g_{2}^2) -181 g_{2}^4}{216 (4 \pi )^6}\\
  \end{array}\right)
\label{MixMatrix}
\end{align}
where index `$1$' corresponds to the operator $\Omega_{ij}\chi^i\chi^j$, and index `$2$' corresponds to $\sigma^2$. The diagrams contributing to
the calculation are listed in Fig.~\ref{diagrams}.

The two eigenvectors of $\gamma^{ab}$ give the two linear combinations of $\sigma^2$ and $\Omega_{ij}\chi^i\chi^j$,
and the eigenvalues $\gamma_{\pm}$ give their anomalous dimensions, so that $\Delta_{\pm}=d-2+\gamma_{\pm}$. We find that one of these combinations is a conformal primary,
and the other one is a descendant of $\sigma$. Indeed, after plugging in the fixed point couplings, we find
$\Delta_-=\Delta_{\sigma}+2$. For instance, in the large $N$ expansion we find the results
\begin{equation}
\begin{aligned}
&\Delta_- = \Delta_{\sigma}+2 = 4+\frac{-40\eps+\frac{104}{3}\eps^2}{N} + \frac{6800\eps- \frac{34190}{3}\eps^2}{N^2}+\ldots\,,\\
&\Delta_+ = 4+\frac{100\eps-\frac{395}{3}\eps^2}{N} +\frac{-49760\eps+\frac{237476}{3}\eps^2}{N^2}+\ldots\,.
\end{aligned}
\end{equation}
Upon sending $N\rightarrow -N$, the dimension $\Delta_+$ can be checked to be in agreement with the available large $N$ results
for the critical exponent $\omega$ in the quartic $O(N)$ theory\footnote{This critical exponent is related to the derivative of the beta function in the $O(N)$ theory with quartic interaction; see \cite{Fei:2014xta} for more details on the comparison to the large $N$ results.}
\cite{Lang:1992zw, Broadhurst:1996ur}.


\subsection{Estimating operator dimensions with Pad\'e approximants}

\begin{figure}[t]
\centering
\includegraphics[width=10cm]{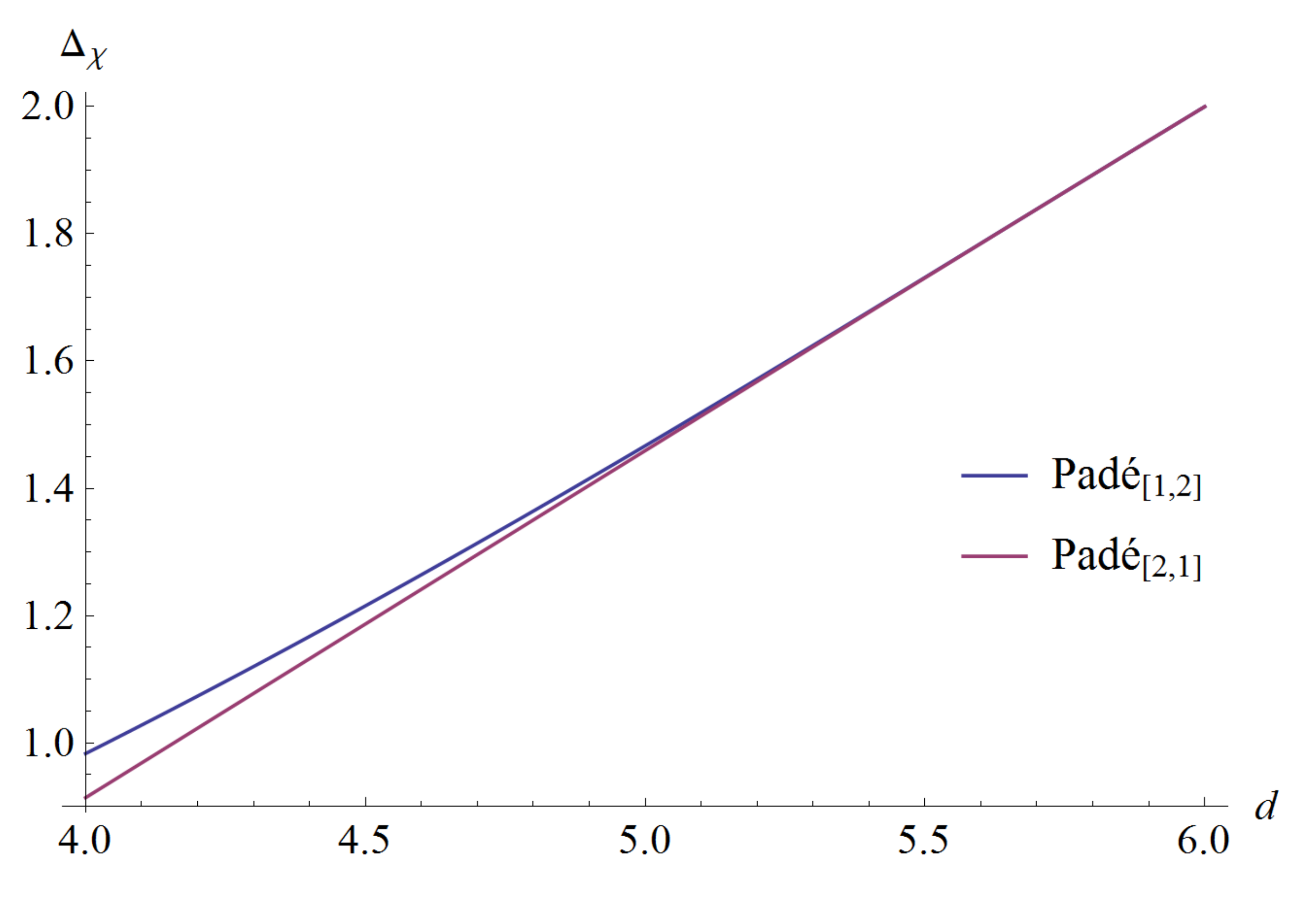}
\caption{Different Pad\'e approximations of $\Delta_{\chi}$ for the $N=2$ model.}
\label{PadeSp2}
\end{figure}

In the previous section we calculated $\epsilon$ expansions for the operator dimensions of the cubic $Sp(N)$ theory in $d=6-\eps$.
We may use the following Pad\'e approximant to estimate the behavior of operator dimensions as we continue $d$:
\begin{equation}
\text{Pad\'e}_{m,n}[d]=\frac{A_0 +A_1 d+ A_2 d^2+\ldots+A_m d^m}{1+B_1 d + B_2 d^2+\ldots+B_n d^n}.
\end{equation}
For $N=2$, by demanding that the $\eps$ expansion have the same behavior as (\ref{N2-dim}), we can fix four coefficients in the Pad\'e approximant. Here we use $\text{Pad\'e}_{1,2}$, and $\text{Pad\'e}_{2,1}$, to obtain the following estimate for the operator dimension in $d=5$:
\begin{align}
\Delta_\chi^{N=2} = \Delta_\sigma^{N=2} \approx
\begin{cases}
1.467 \qquad \text{with Pad\'e}_{1,2}\\
1.459 \qquad \text{with Pad\'e}_{2,1}\ .
\end{cases}
\end{align}
As expected, this is below the unitarity bound in $d=5$, which is $\Delta=\frac{3}{2}$. The plots of different Pad\'e approximants are shown in Fig.~\ref{PadeSp2}.
For the next primary operator, whose scaling dimension is given in (\ref{delp}), using the $\text{Pad\'e}_{2,1}$ approximant we estimate
$\Delta_+\approx 3.35$ in $d=5$.
In four dimensions we estimate $\Delta_+\approx 2.7$; this suggests that the $N=2$ model does not have a free field description near
$d=4$.

In the $Sp(4)$ case, we may assume for $4<d<6$ that the IR fixed point of the cubic theory (\ref{cubicsymplectic})
is equivalent to the UV fixed point of the quartic theory (\ref{symplectic}).\footnote{This equivalence holds for $Sp(N)$ models with sufficiently large $N$, but it is not completely clear if it applies for $N=4$.}
If so, then not only do we know the $6-\eps$ expansion of $\Delta_\chi$ to $\mathcal{O}(\eps^3)$, we can also use the quartic theory result in $d=4+\eps$, which is known to $\mathcal{O}(\eps^5)$:
\begin{align}
\Delta_\chi =\begin{cases}
2-0.529827 \eps -0.0126197 \eps^2 + 0.0157244 \eps^3 \qquad\qquad\qquad\qquad\qquad\;\; \text{in } d=6-\eps\\
1+ 0.5 \eps -0.03125 \eps^2 + 0.015625 \eps^3 - 0.0730951 \eps^4 + 0.0195503 \eps^5 \qquad \text{in } d=4+\eps
\end{cases}
\end{align}
Together, these expansions allow us to determine ten coefficients of the Pad\'e approximant. Disregarding the approximants that have a pole, we obtain the estimate $\Delta_\chi^{N=4} \approx 1.478$ in $d=5$.

\subsection{Sphere free energies}

It is also interesting to compute the sphere free energy at the IR fixed point in $d=6-\epsilon$. The leading order term in the $\epsilon$ expansion for
the corresponding $O(N)$ models was computed in \cite{Giombi:2014xxa}. Sending $N\rightarrow -N$, we find the following result for
$\tilde F = \sin(\frac{\pi d}{2})\log Z_{S^d}$ in the cubic $Sp(N)$ models
\begin{equation}
\tilde F_{\textrm{IR}} = \tilde F_{\textrm{UV}} -\frac{\pi}{17280}\frac{(g_2^*)^2-3N(g_1^*)^2}{(4\pi)^3}\eps+\mathcal{O}(\eps^3),
\end{equation}
where $\tilde F_{\textrm{UV}} = (1-N)\tilde F_s$, and $\tilde F_s$ the value corresponding to a free conformal scalar. Plugging in the explicit solutions
for the fixed point couplings $g_1^*$, $g_2^*$, it is straightforward to verify that, for all $N\ge 2$, we have
\begin{equation}
\tilde F_{\textrm{UV}} > \tilde F_{\textrm{IR}}\,.
\label{fthm}
\end{equation}
For instance, for $N=2$ we find
\begin{equation}
\label{OSpF}
\tilde F_{\textrm{IR}} =-\tilde F_s  -\frac{\pi}{43200}\eps^2+\mathcal{O}(\eps^3)\,.
\end{equation}
Note that for the Fisher model \cite{Fisher:1978pf} of the Lee-Yang edge singularity, corresponding to $N=0$ and imaginary $g_2^*$,
the inequality (\ref{fthm}) does not hold. For general non-unitary theories the inequality does not have to hold; remarkably, it does hold for the
$Sp(N)$ models with positive $N$.

Similarly, using the results of \cite{Giombi:2014xxa} for the quartic $O(N)$ theory in $d=4-\epsilon$, we can compute the sphere free energy at the IR fixed
point of the model (\ref{symplectic}) in $d=4-\epsilon$. We find
\begin{equation}
\tilde F_{\textrm{quartic}} = \tilde F_{\textrm{free}} - \frac{\pi}{576}\frac{N(N-2)}{(N-8)^2}\eps^3+\mathcal{O}(\eps^4)\ ,
\end{equation}
where $\tilde F_{\textrm{free}} = -N \tilde F_s$.
We observe that $\tilde F_{\textrm{free}} > \tilde F_{\textrm{quartic}}$ is satisfied for all $N>2$.
\footnote{For $\eps <0$ the interacting theory has a UV fixed point. So, in this case
$\tilde F_{\textrm{free}} < \tilde F_{\textrm{quartic}}$, again in agreement with the conjectured $\tilde F$ theorem.}

The case $N=8$ is special and needs to be treated separately. Here the one-loop
term in the beta function vanishes, and we have
\be
\beta_\lambda=-\eps \lambda +{15\over 32 \pi^4} \lambda^3+ \mathcal{O}(\lambda^4).
\ee
Therefore, at the IR fixed point $\lambda_*^2=  {32 \pi^4 \over 15} \eps + \mathcal{O}(\eps^{3/2})$, and we get
\begin{equation}
\tilde F_{\textrm{quartic}} = \tilde F_{\textrm{free}} - \frac{\pi}{360}\eps^2+\ldots.
\end{equation}




\section{Symmetry enhancement for $N=2$}
\label{symenh}

Let us write the cubic $Sp(2)$ model in terms of a real scalar $\sigma$ and a single complex anti-commuting fermion $\theta$:
\be
\label{cubicOSp}
S=\int d^d x \bigg (
\partial_\mu \theta \partial^\mu \bar \theta + \frac{1}{2}\left(\partial_{\mu}\sigma\right)^2 +g_1 \sigma \theta \bar \theta + \frac{1}{6}g_2 \sigma^3
\bigg )
\ .\ee
For $g_2=2 g_1$, i.e. the fixed point relation (\ref{N2-fixed}), this action possesses a fermionic symmetry with a complex anti-commuting scalar parameter $\alpha$
\be \delta \theta = \sigma \alpha\ , \quad
\delta \bar \theta = \sigma \bar \alpha\ ,\quad
\delta \sigma = - \alpha \bar \theta + \bar \alpha \theta\ .
\ee
As a consequence of this symmetry, the scaling dimensions of $\sigma$ and $\theta$ are equal, as seen explicitly in eq. (\ref{N2-dim}).
This complex fermionic symmetry enhances the $Sp(2)$ to $OSp(1|2)$, which is the smallest supergroup.
The full set of supergroup generators can be given in the form
\be
\begin{aligned}
& Q^{+} = \frac{1}{2} \left (\sigma {\partial\over \partial \bar\theta} +\theta {\partial \over \partial\sigma} \right )\,,\qquad  Q^{-} = \frac{1}{2} \left (\sigma {\partial\over \partial \theta} -\bar \theta {\partial \over \partial\sigma} \right ), \\
& J^{+} =\theta {\partial \over \partial\bar\theta }\,,\qquad
J^{-} =\bar\theta {\partial \over \partial \theta}\,,\qquad
J^3= \frac{1}{2} \left (\theta {\partial\over \partial \theta} -\bar \theta {\partial \over \partial \bar \theta} \right )\ , \\
\end{aligned}
\ee
and it is not hard to check that they satisfy the algebra of $OSp(1|2)$:
\begin{align}
\left[J^3, J^{\pm} \right] &= \pm J^{\pm}, \quad \left[J^+, J^- \right]=2J^3 \\
\left[J^3, Q^{\pm} \right] &= \pm \frac{1}{2}Q^{\pm}, \quad \left[J^{\pm}, Q^{\mp} \right] = -Q^{\pm} \\
\{Q^{\pm}, Q^{\pm} \}&=\pm\frac{1}{2}J^{\pm}, \quad \{Q^{+}, Q^{-} \}=\frac{1}{2}J^3
\end{align}

We expect the operators of the theory to form representations of $OSp(1|2)$. For example,
let us study the mixing of the $Sp(2)$ singlet operators $\sigma^2$ and $\theta\bar\theta$. Setting $N=2$ in (\ref{MixMatrix}), we find the scaling dimensions of the two eigenstates:
\begin{equation}
\Delta_+ = 4-\frac{2}{3}\eps + \frac{1}{30}\eps^2+\mathcal{O}(\eps^3)\,,\qquad
\Delta_- = 4-\frac{8}{15}\eps - \frac{7}{450}\eps^2+\mathcal{O}(\eps^3).
\label{delplus}
\end{equation}
The first of these dimensions corresponds to the conformal primary operator
\begin{equation}
O_+ =\sigma^2+2\theta\bar{\theta},
\end{equation}
which is invariant under all the $OSp(1|2)$ generators.
The second corresponds to $O_- =\sigma^2 + \theta \bar{\theta}$, which is a conformal descendant because at the fixed point it is proportional to
$\partial_\mu \partial^\mu \sigma$ by equations of motion. Indeed, we find $\Delta_-=\Delta_{\sigma}+2$.

Continuation of the results for the cubic model with $OSp(1|2)$ global symmetry to finite $\epsilon$ points to the existence of such interacting CFTs in integer dimensions
below 6. 
In \cite{Caracciolo:2004hz} it was argued that
the $q\rightarrow 0$ limit of the $q$-state Potts model 
is described by the $OSp(1|2)$ sigma model \cite{Read:2001pz,Saleur:2003zm}.
There is evidence that the upper critical dimension of the spanning forest model is $6$ \cite{2007PhRvL..98c0602D}, and the $6-\epsilon$ expansions of
the critical indices are \cite{deAlcantaraBonfim:1980pe,2007PhRvL..98c0602D}
\begin{align}
&\eta = - \frac{1}{15}\epsilon - \frac{7}{225} \epsilon^2-\frac{269 - 702 \zeta(3)}{16875} \epsilon^3+\mathcal{O}(\eps^4)\,, \\
&\nu^{-1} = 2-\frac{1}{3}\eps - \frac{1}{30}\eps^2-\frac{173 - 864 \zeta(3)}{27000} \epsilon^3+\mathcal{O}(\eps^4)\ .
\label{critexp}
\end{align}
Remarkably, the operator dimensions we have calculated (\ref{N2-dim}) and (\ref{delplus}) agree with these expansions upon the standard identifications
\begin{equation}
\label{delp}
\Delta_\chi = \frac{d}{2}-1 + \frac{\eta}{2}\ , \qquad \Delta_+= d- \nu^{-1}\ .
\end{equation}
Similarly, we can match the $6-\eps$ expansions of the sphere free energies. For the $q$-state Potts model it is not hard to show that
\begin{equation}
\tilde F_{q} = (q-1)\tilde F_{s} +\frac{\pi (q-1)(q-2)}{8640 (3 q-10)}\eps^2+\mathcal{O}(\eps^3)\,,
\end{equation}
and for $q=0$ this matches the $\tilde F$ of our $OSp(1|2)$ model, (\ref{OSpF}).
These results provide strong evidence that the $OSp(1|2)$ symmetric IR fixed point of the cubic theory (\ref{cubicOSp}) describes the
$q\rightarrow 0$ limit of the $q$-state Potts model.

Numerical simulations of the spanning-forest model \cite{2007PhRvL..98c0602D}, which is equivalent to the $q\rightarrow 0$ limit of the ferromagnetic $q$-state Potts model, indicate the
existence of second-order phase transitions in dimensions 3, 4 and 5. The critical exponents found in \cite{2007PhRvL..98c0602D}
are in good agreement with the Pade extrapolations of $6-\eps$ expansions exhibited in section 2.1.
This provides additional evidence for the existence of the critical theories with $OSp(1|2)$ symmetry.
It would be of further interest to
find the critical statistical models that are described by the $Sp(N)$ invariant theories with $N>2$.

\section*{Acknowledgments}

We thank D. Anninos, D. Harlow and J. Maldacena, and especially S. Caracciolo and G. Parisi, for useful discussions.
The work of LF and SG was supported in part by the US NSF under Grant No.~PHY-1318681.
The work of IRK and GT was supported in part by the US NSF under Grant No.~PHY-1314198.


\bibliographystyle{ssg}
\bibliography{Sp}

\end{document}